%% file: LimitsOfFixedLength.tex
\documentclass[journal]{IEEEtran}
 \usepackage{amsmath}
 \usepackage{amsfonts}
 \usepackage{amssymb}
 \usepackage{graphicx}
 \usepackage{xcolor}
 \usepackage{pgfplots}
 \usepackage{todonotes}
 \usepackage{mathtools}
 \input{preamble}

 \author{\IEEEauthorblockN{Patrick Schulte, Bernhard C. Geiger}\\
 \IEEEauthorblockA{Institute for Communications Engineering, Technical University of Munich, 80333, Munich, Germany\\
 Emails: $\{$patrick.schulte,bernhard.geiger$\}$@tum.de}%
 }
 
 \title{Divergence Scaling of Fixed-Length, Binary-Output, One-to-One Distribution Matching}

 \DeclareMathOperator{\supp}{supp}

 \newcommand{\Pletter}{\ensuremath{P_{\rv{\bar{A}},\mset{C}}}}
 \renewcommand{\Pletter}{\ensuremath{\hat{P}_{\mset{C}}}} 
 \newcommand{\Pletterk}{\ensuremath{\hat{P}_{\mset{C}_k}}} 
 \newcommand{\pletter}{\ensuremath{p_{\mset{C}}}} 
 \newcommand{\pletterk}{\ensuremath{p_{\mset{C}_k}}}

 \newcommand{\kopt}{\ensuremath{\hat{k}}}
 \newcommand{\qopt}{\ensuremath{\hat{q}}}
 \newcommand{\popt}{\ensuremath{p_{\Copt}}}
 \newcommand{\Copt}{\ensuremath{\mset{C}_{\kopt}}}
 \newcommand{\unif}[1]{U_{#1}}

\begin{document}

\maketitle
\begin{abstract}
	Distribution matching is the process of invertibly mapping a uniformly distributed input sequence onto sequences that approximate the output of a desired discrete memoryless source.
	The special case of a binary output alphabet and one-to-one mapping is studied.
	A fixed-length distribution matcher is proposed that is optimal in the sense of minimizing the unnormalized informational divergence between its output distribution and a binary memoryless target distribution.
	Upper and lower bounds on the unnormalized divergence are computed that increase logarithmically in the output block length $n$.
	It follows that a recently proposed constant composition distribution matcher performs within a constant gap of the minimal achievable informational divergence.
\end{abstract}
	
\section{Introduction and Preliminaries}
A distribution matcher (DM) transforms sequences of independent, uniformly distributed symbols into sequences that approximate a discrete memoryless source (DMS) with a target distribution (see Fig.~1). A dematcher performs the inverse operation and recovers the input symbols from the output sequence.

Optimal variable-to-fixed- and fixed-to-variable-length DMs are proposed in~\cite{kschischang1993optimal,Ungerboeck2002,bocherer2011matching,amjad2013fixed}.
The codebooks of these DMs must be generated offline and stored.
Since this is infeasible for large codeword lengths, % that are necessary to achieve the maximum rate.
schemes were proposed that use arithmetic coding to calculate the codebook online~\cite{Cai2007,bocherer2013arithmetic}.
All these approaches have either variable input or variable output lengths,
which can lead to varying transmission rate, large buffer sizes, error propagation, and synchronization problems \cite[Sec.~I]{kschischang1993optimal}.

Fixed-to-fixed-length DMs do not suffer from these problems.
The authors of \cite[Sec.~4.8]{amjad2013algorithms} and~\cite{mondelli2014achieve} therefore suggest to use block codes to build a fixed-to-fixed-length DM. 
However, these schemes include many-to-one mappings and hence cannot always recover the input sequence without error.

To overcome these problems, in~\cite{schulte2015constant} a fixed-to-fixed-length, binary-output, one-to-one DM was proposed with a codebook being a subset of a \emph{type set}.
The normalized divergence between the output distribution of this constant composition distribution matcher (CCDM) and the target distribution vanishes as the output block length tends to infinity.
This property is important for energy efficient communication schemes such as the scheme proposed in~\cite{bocherer2011operating,mondelli2014achieve} that achieves the capacity of arbitrary discrete memoryless channels (DMCs) \cite{mondelli2014achieve} or the scheme proposed in \cite{bocherer2015bandwidth} that achieves the capacity of the additive white Gaussian noise channel.

Another application of DMs is stealth communication \cite{hou2014effective}, in which the adversary should not learn whether a transmission occurs or not. To achieve stealth, the \emph{unnormalized} divergence between the received signal and an ``idle signal'' of the channel (e.g. white Gaussian noise) should approach zero. 
For example, in~\cite[Lemma~1]{bocherer2011matching} is was shown that their variable-to-fixed-length DM achieves an unnormalized divergence that is upper bounded by a constant, independently of the output length.

To the best of the authors' knowledge, the performance of fixed-to-fixed-length DMs has not yet been evaluated in terms of unnormalized divergence.
This work aims to fill this gap. In particular, we show that even for the optimal fixed-to-fixed-length, binary-output, one-to-one DM, the unnormalized divergence increases at least logarithmically in the output block length (Section~\ref{sec:optimalConstruction}).
Our results thus suggest that practical one-to-one, fixed-to-fixed-length DMs cannot provide stealth for all possible channels (e.g., the identity channel). %Future work shall investigate this issue further. 

Furthermore, we show in Section~\ref{sec:ccdm} that the unnormalized divergence of CCDM increases also logarithmically in the output block length. Thus, CCDM achieves the minimum unnormalized divergence within a constant gap.
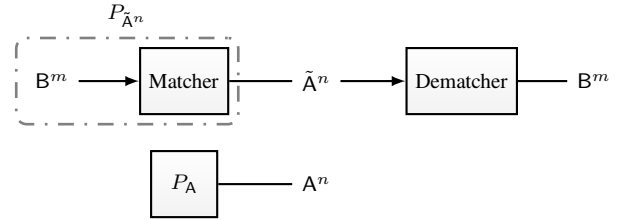
\begin{figure}[t]
	\centering
	\input{diagrams/memoryless_Source}
	\caption{Matching input block $\rv{B}^m = (\rv{B}_1,\ldots,\rv{B}_m)$ to output symbols $\rv{\tilde A}^n = (\rv{\tilde A}_1,\ldots,\rv{\tilde A}_n)$ and reconstructing
		the original sequence at the dematcher. The DM should emulate a DMS with distribution $\pmf{A}$.}
	\label{fig:DMS model}
\end{figure}

\section{Preliminaries}\label{sec:preliminaries}
We denote a random variable as $\rv{A}$, its alphabet as $\mset{A}$, and its distribution as $\pmf{A}$. For binary alphabets $\mset{A}=\{0,1\}$ we sometimes write $p = \pmf{A}(1) = 1- \pmf{A}(0)$. The uniform distribution on the set $\mset{A}$ is denoted as $\unif{\mset{A}}$, i.e., for every $a\in\mset{A}$, we have $\unif{\mset{A}}(a)=1/|\mset{A}|$.

We denote a length-$n$ random string or sequence as $\rv{{A}}^n := (\rv{{A}}_1,\!\ldots\!,\rv{{A}}_n)$. If its entries are independent and identically distributed (iid), we have, for every realization $a^n:=(a_1,\!\ldots\!,a_n)\in\mset{A}^n$,
	\begin{equation}  	
	\pmfn{A}{n}(a^n)=\pmf{A}^n(a^n) = \prod_{i=1}^n \pmf{A}(a_i).  
	\end{equation}

The empirical distribution of the sequence $a^n\in \mset{A}^n$ is
    \begin{equation}
    \label{eq:empiricalDist}  	\hat{P}_{{a^n}}(\alpha) = \frac{n_\alpha(a^n)}{n}
    \end{equation}
where $n_\alpha(a^n) = \left| \left\lbrace i: a_i = \alpha \right\rbrace \right|$ is the number of times the symbol $\alpha\in \mset{A}$ appears in $a^n$.
The authors of \cite[Sec.~2.1]{csiszar2004information} call $\hat{P}_{{c}}$ the \emph{type} of the sequence ${c}$.  	
An $n$-type is a type based on a length-$n$ sequence.  	
Note that the $n$-types partition the alphabet $\mset{A}^n$ into equivalence classes, called the \emph{type sets}.

If $\mset{A} = \lbrace 0,1\rbrace$, we call $n_1(a^n)$ the \emph{weight} of $a^n$. 
In this case we write $\mset{C}_k$ for the union of all type sets up to weight $k$. We have
\begin{equation*}
|\mset{C}_k| = \sum_{i = 0}^{k} \binom{n}{i}.
\end{equation*}

The entropy of a discrete random variable $\rv{A}$ with alphabet $\mset{A}$ and distribution $\pmf{A}$ is  	
\begin{equation}
\entrp{\pmf{A}} = \sum_{a \in \supp(\pmf{A})} -\pmf{A}(a) \log_2 {\pmf{A}(a)}
\end{equation}  	
where $\supp(\pmf{A}) \subseteq \mset{A}$ is the support of $\pmf{A}$. 
We sometimes write $\entrp{p}$ for binary entropies. 	

The informational divergence of two distributions on $\mset{A}$ is  	\begin{equation}  	
\diverg{\pmf{\hat{A}}}{\pmf{A}} = \sum_{a \in \supp(\pmf{\hat{A}})} \pmf{\hat{A}}(a) \log_2 \frac{\pmf{\hat{A}}(a)}{\pmf{A}(a)}.
\end{equation}  	
We sometimes write $\diverg{\hat{p}}{p}$ for the informational divergence of two distributions on $\mset{A}=\{0,1\}$ with $\pmf{\hat{A}}(1)=\hat{p}$ and $\pmf{A}(1)=p$.
With this notation, we have the following result.  	
\begin{lemma}[Bound on Entropy Difference]\label{lem:asymptoticEntropyB}
	Let $0<p<1$ and $0<p-\epsilon<p$. Then  		
	\begin{align}  		 
	\entrp{p} - \entrp{p-\epsilon} &\leq \epsilon\log_2\frac{1-p+\epsilon}{p-\epsilon}.  		
	\end{align}  	
\end{lemma}

\begin{IEEEproof}
  The bound follows from
  \begin{equation}
    \entrp{p} - \entrp{p-\epsilon} = \epsilon\log_2\frac{1-p+\epsilon}{p-\epsilon} - \diverg{p}{p-\epsilon}
  \end{equation}
  and the non-negativity of informational divergence.
\end{IEEEproof}

Our results rely on properties of binomial coefficients. We state the relevant results here.

\begin{lemma}[Bounds on binomial coefficient]\label{Lem:ashBoundBinomial}
	If $0 < p < 1$ and $np$ is integer, then
	\begin{equation}
	\frac{2^{n \entrp{p}}}{\sqrt{8np(1-p)}}  \leq \binom{n}{np} \leq \frac{2^{n \entrp{p}}}{\sqrt{2\pi np(1-p)}} .
	\end{equation}
\end{lemma}
\begin{IEEEproof}
	  The proof follows by applying Stirling's approximation twice, see \cite[Ch.~4.7]{ash1965information}.
\end{IEEEproof}

\begin{lemma}[Bounds on partial sums of binomial coefficients~\cite{bahadur1960approximations}]\label{lem:BoundsSumBinomial} If $0 \leq p < 1/2$ and $np$ is integer, then
	\begin{equation}
	\binom{n}{np} 
	\alpha\beta\leq \sum_{i=0}^{np} \binom{n}{i}\leq \binom{n}{np}\alpha 
	\end{equation}
	with
	\begin{equation}
	\alpha = \frac{1-p+1/n}{1-2p+1/n}
	\end{equation}
	and
	\begin{equation}
	\beta = \frac{n(1-2p)^2}{1+n(1-2p)^2}.
	\end{equation}	
\end{lemma}
\begin{lemma}[Sum of binomial coefficients weighted with distance from the center~{\cite[p.~166]{graham1989concrete}}]\label{lem:SumCenterweightedBinomials} For every positive integer $n$ and  non-negative integer $k$, $k\leq n$, we have
	  \begin{align*}
	  \sum_{i = 0}^{k} \binom{n}{i}\left(\frac{n}{2} -i\right) = \frac{k+1}{2}\binom{n}{k+1}.
	  \end{align*}
\end{lemma}

\section{Fixed-Length Distribution Matching}\label{sec:DistributionMatching}
Since variable-length DMs suffer from error propagation and large buffer sizes \cite[Sec.~I]{kschischang1993optimal}, we focus on fixed-length DMs. A fixed-length DM is a mapping $f{:}\ \mset{B}^m \to \mset{A}^n$ that maps a length-$m$ input block $\rv{B}^m$ onto a length-$n$ output block $\rv{\tilde{A}}^n$ with distribution $\pmfn{\tilde{A}}{n}$ that should mimic the output of a DMS with distribution $\pmf{A}$ on $\mset{A}$ (see Fig.~\ref{fig:DMS model}). The mapping's image $\mset{C}= f(\mset{B}^m)$ is called the \emph{codebook} and each element in the codebook is a \emph{codeword}. In this work, we focus on mappings $f$ that are one-to-one. However, it may be interesting to also consider invertible one-to-many and invertible random mappings.

Application areas for DMs are energy efficient communication and stealth communication.
In both areas, the informational divergence between the output distribution of the DM and the target distribution plays a fundamental role.
		
For energy efficient communication, suppose that $\pmf{A}$ is the capacity-achieving input distribution of a DMC with capacity $C$. Let $\rv{\tilde{Y}}^n$ be the channel output for an input $\rv{\tilde{A}}^n$. If $\mathbb{I}(\rv{\tilde{A}}^n;\rv{\tilde{Y}}^n)$ is the mutual information between the input and the output sequence, then it can be shown that~\cite[eq.~(23)]{bocherer2011matching}
\begin{equation}
	C - \frac{\diverg{\pmfn{\tilde{A}}{n}}{\pmf{A}^n}}{n}\leq \frac{\mathbb{I}(\rv{\tilde{A}}^n;\rv{\tilde{Y}}^n)}{n} \leq C.
\end{equation}
Hence, a small \emph{normalized} divergence guarantees a mutual information close to capacity.
	
For stealth communication, suppose that an adversary wants to detect a transmission over a DMC, i.e., the activity rather than the content. Suppose that $\pmf{Y}$ is the distribution the adversary expects to observe at the channel output when \emph{no} transmission occurs, and suppose that $\pmf{A}$ is a distribution for which the channel responds with exactly this output distribution. Let again $\rv{\tilde{Y}}^n$ be the channel output for an input $\rv{\tilde{A}}^n$. In~\cite[Sec.~IV \& Lemma~1]{hou2014effective}, the authors showed that if
\begin{equation}
  \diverg{\pmfn{\tilde{Y}}{n}}{\pmf{Y}^n} \to 0
\end{equation}	
as $n\to \infty$, the best an adversary can do is to guess without observing $\rv{\tilde{Y}}^n$.
Since, by the data processing inequality, $\diverg{\pmfn{\tilde{A}}{n}}{\pmf{A}^n}\ge\diverg{\pmfn{\tilde{Y}}{n}}{\pmf{Y}^n}$, zero \emph{unnormalized} divergence guarantees \emph{stealth}.
Note, however, that this is only a sufficient condition: Even if $\diverg{\pmfn{\tilde{A}}{n}}{\pmf{A}^n}>0$, one can obtain $\diverg{\pmfn{Y}{n}}{\pmf{Y}^n}\to 0$ depending on the DMC\footnote{If the DMC is completely noisy, one gets $\diverg{\pmfn{Y}{n}}{\pmf{Y}^n}= 0$ for every $\pmfn{\tilde{A}}{n}$. In contrast, if the the DMC is noiseless, then the condition $\diverg{\pmfn{\tilde{A}}{n}}{\pmf{A}^n}\to 0$ becomes necessary for stealth.}.

Throughout this work we assume that $\rv{B}^m$ is a sequence of independent, uniformly distributed random variables $\rv{B}_i$, hence $\rv{B}^m$ is uniformly distributed on $\mset{B}^m$. Since $f$ is one-to-one, we have $P_{f(\rv{B}^m)}=\unif{\mset{C}}$, and the informational divergence can be calculated by
\begin{align}\label{eq:basicDivergence}
\diverg{\unif{\mset{C}}}{P_A^n} =\sum_{a^n \in \mset{C}} \frac{1}{|\mset{C}|} \log_2 \frac{\frac{1}{|\mset{C}|}}{\pmf{A}^n(a^n)}.
\end{align}
We define the \emph{letter distribution} $\Pletter$ of the codebook as
\begin{equation}\label{eq:def_letterdistrib}
\Pletter(\alpha) = \frac{1}{|\mset{C}|} \sum_{a^n \in \mset{C}} \hat{P}_{a^n}(\alpha)
\end{equation}
which corresponds to the probability of drawing a letter $\alpha\in\mset{A}$ from the whole codebook.
By manipulating the divergence expression we get
\begin{align}
&\diverg{\unif{\mset{C}}}{\pmf{A}^n} \nonumber \\
&= \sum_{a^n \in \mset{C}} \frac{1}{|\mset{C}|} \log_2 \frac{\frac{1}{|\mset{C}|}}{\Pletter^n(a^n)}
+\sum_{a^n \in \mset{C}} \frac{1}{|\mset{C}|} \log_2 \frac{\Pletter^n(a^n)}{\pmf{A}^n(a^n)} \nonumber\\
&= \diverg{\unif{\mset{C}}}{\Pletter^n} 
+ \sum_{a^n \in \mset{C}} \frac{1}{|\mset{C}|} \sum_{\alpha\in\supp(\Pletter)}n_\alpha(a^n)\log_2 \frac{\Pletter(\alpha)}{\pmf{A}(\alpha)}\nonumber\\
%&= \diverg{\unif{\mset{C}}}{\Pletter^n} 
%+ n \sum_{\alpha\in\mset{A}} \sum_{a^n \in \mset{C}} \frac{1}{|\mset{C}|} %\frac{n_\alpha(a^n)}{n}\log_2 %\frac{\Pletter(\alpha)}{\pmf{A}(\alpha)}\nonumber\\
&\stackrel{(a)}{=}\diverg{\unif{\mset{C}}}{\Pletter^n} +n\diverg{\Pletter}{\pmf{A}}\label{eq:sumdiv},
\end{align}
where $(a)$ follows from~\eqref{eq:empiricalDist} and~\eqref{eq:def_letterdistrib}. Similarly, we have 
\begin{equation}\label{eq:firstterm}
\diverg{\unif{\mset{C}}}{\Pletter^n} = -\log_2|\mset{C}| + n\entrp{\Pletter}.
\end{equation}
Note that \eqref{eq:firstterm} depends only on the selected codebook and is independent of the target distribution.

We consider \emph{binary-output} DMs, i.e., $\mset{A}=\{0,1\}$. We write $p=\pmf{A}(1)=1-\pmf{A}(0)$ and $\pletter = \Pletter(1) = 1- \Pletter(0)$. For~\eqref{eq:sumdiv} we have
\begin{equation}\label{eq:binaryCost}
\diverg{\unif{\mset{C}}}{\pmf{A}^n} = -\log_2|\mset{C}| + n\entrp{\pletter} +n\diverg{\pletter}{p}.
\end{equation}
We suppose that $0<p<\frac{1}{2}$ because this encompasses all interesting cases.

\section{Analysis of CCDM}\label{sec:ccdm}
In \cite{schulte2015constant}, a fixed-length, binary-output, one-to-one DM was presented for which all codewords in the codebook $\mset{C}$ have the same composition or type,
i.e., $\hat{P}_{a^n}$ is the same for all $a^n\in\mset{C}\subset\{0,1\}^n$. The DM is called a constant composition distribution matcher~(CCDM). For a given output sequence length $n$, we wish to choose the type $\hat{P}_\mset{C}$ that approximates the binary target distribution $ \pmf{A}(1) = 1- \pmf{A}(0) = p $ optimally in the sense of minimizing $\mathbb{D}(\Pletter\Vert\pmf{A})$.
This optimal $\Pletter$ is called the optimal $n$-type approximation of $\pmf{A}$, cf.~\cite{bocherer2014optimal}.

For CCDMs we can find reasonably tight bounds for unnormalized divergence.
In particular, there exists a constant $\kappa$ such that $\mathbb{D}(\Pletter\Vert\pmf{A}) < \kappa/n^2$~\cite[Prop.~4]{bocherer2014optimal}.
Moreover, with~\eqref{eq:firstterm} we have
\begin{equation}
  \diverg{\unif{\mset{C}}}{\Pletter^n} 
  = -\log_2|\mset{C}| + n\entrp{\pletter} = \log_2\frac{2^{n\entrp{\pletter}}}{|\mset{C}|}.
\end{equation}
This divergence is minimized by choosing the largest possible codebook,
i.e., by selecting as many codewords from the chosen type set as possible.
If the input symbols are $B$-ary, i.e., $|\mset{B}|=B$, the codebook size is a power of $B$.
Moreover there exists an integer $m$ such that $B^m \leq \binom{n}{n\pletter} < B^{m+1}$.
Thus, if we choose the codebook such that $|\mset{C}| = B^m$ we get $|\mset{C}| > \binom{n}{n\pletter}/B$ and with \eqref{eq:firstterm} and Lemma \ref{Lem:ashBoundBinomial} we obtain
\begin{align*}
	\diverg{\unif{\mset{C}}}{\Pletter^n}  &< \log_2\frac{2^{n\entrp{\pletter}}}{\binom{n}{n\pletter}/B}\\
%  	 	&\leq \frac12\log_2 (\bg{32}n \pletter(1-\pletter)) \\
	&\leq \frac12 \log_2n + \frac12\log_2(8B\pletter(1-\pletter)).
\end{align*}
Since $\pletter\to p$ for large $n$, it follows that $\diverg{\unif{\mset{C}}}{\pmf{A}^n}$ grows at most logarithmically in $n$. Employing the upper bound from Lemma \ref{Lem:ashBoundBinomial} together with $|\mset{C}|\le\binom{n}{n\pletter}$ yields 
\begin{equation*}
	\diverg{\unif{\mset{C}}}{\Pletter^n} \ge \frac12 \log_2 n + \frac12\log_2(2\pi \pletter(1-\pletter)) 
\end{equation*}	
and hence $\diverg{\unif{\mset{C}}}{\pmf{A}^n}$ grows logarithmically in $n$.

Clearly, the above CCDM may be used in energy efficient transmission schemes for DMCs as normalized divergence approaches zero.
However, its unnormalized divergence does not vanish.
One may claim that this is a consequence of the following two restrictions:
1) all codewords in the codebook were chosen from the same type set,
and 2) the codebook size is restricted to be a power of $B$.
In the following section we drop both restrictions and present an optimal, albeit impractical, codebook construction.
We show that $\diverg{\unif{\mset{C}}}{\pmf{A}^n}$ still grows at least logarithmically in $n$. Hence, this growing divergence is not a consequence of the two restrictions mentioned above, but rather a fundamental limit of fixed-length, binary-output, one-to-one DMs. As a side result, one can see that
CCDM achieves the unnormalized divergence of the optimal scheme within a constant gap.

\section{Optimal Codebook Construction}\label{sec:optimalConstruction}
In \cite{amjad2013algorithms} a codebook construction that is optimal for a fixed-length, binary-output, one-to-one DM is proposed.
For a given codebook size $|\mset{C}|$, the divergence
\begin{align}
  \diverg{\unif{\mset{C}}}{P_A^n}
  =\sum_{a^n \in \mset{C}} \frac{1}{|\mset{C}|} \log_2 \frac{\frac{1}{|\mset{C}|}}{\pmf{A}^n(a^n)}
\end{align}
is small if we choose $\mset{C}$ to be an ensemble of codewords that are most likely according to the target distribution $\pmf{A}^n$. Since $p<1/2$, $\pmf{A}^n(a^n)$ is monotonically decreasing in the weight of $a^n$.
Consequently, the codebook construction  should include the all-zero codeword.
Next, codewords with a single one and $n-1$ zeros are included, and so on.
It follows that $\pletter$ grows monotonically in the codebook size $|\mset{C}|$.
It remains to determine the optimal codebook size $|\mset{C}|$, which can be done by a line search \cite{amjad2013algorithms}.

We now characterize the optimal codebook.
In particular, it turns out that the optimal binary codebook is, for some $k$, the union of the $k$ type sets with the lowest weight.
This result seems surprising, as for large $n$ the type sets grow exponentially for a $k$ growing linearly with $n$ (cf.~Lemma~\ref{Lem:ashBoundBinomial}).
We would have guessed that such codebooks are suboptimal in terms of $\diverg{\unif{\mset{C}}}{P_A^n}$.
The following lemma shows that this intuition is wrong.
 	
\begin{lemma}Minimum divergence codebooks are a union of type sets. More precisely, the minimum divergence codebook $\mset{C}$ satisfies $\mset{C} = \mset{C}_k$, for some $k$.
\end{lemma}
\begin{IEEEproof}
	Suppose $\mset{C}$ consists of all type sets with weight at most $k$ and $0\leq\ell \leq \binom{n}{k+1}$ codewords of the type set with weight $k+1$.
	We have
	\begin{equation}\label{eq:continuous_CB}
	|\mset{C}| = |\mset{C}_k| + \ell
	\end{equation}
	and $|\mset{C}_k| \leq |\mset{C}| \leq |\mset{C}_{k+1}|$.
	Using \eqref{eq:continuous_CB} and \eqref{eq:firstterm},
	we obtain
	\begin{align*}
	\pletter =& \frac{\sum_{i=0}^{k} \binom{n}{i} i + \ell(k+1)}{n (|\mset{C}_k| + \ell)}\\
	\diverg{U_\mset{C}}{\pmf{A}^n} =& -\log_2\left(|\mset{C}_k| + \ell\right)  + a\frac{\sum\limits_{i=0}^{k} \binom{n}{i} i + \ell(k+1)}{ \left(|\mset{C}_k| + \ell\right)}+c
	\end{align*}
	with positive constants $a =\log_2((1-p)/p)$ and $ c = -n\log_2(1-p)$. Taking $\ell$ as a continuous variable, the first and second derivatives with respect to $\ell$ are
	\begin{align}
	\frac{\partial}{\partial\ell} \diverg{U_\mset{C}}{\pmf{A}^n} &=
	-\frac{1}{\ln(2)(|\mset{C}_k| + \ell)}
	+ a\frac{\sum\limits_{i=0}^{k}\binom{n}{i}(k+1-i)}{ \left(|\mset{C}_k| + \ell\right)^2}\label{eq:firstderivative}\\
	\frac{\partial^2}{\partial\ell^2} \diverg{U_\mset{C}}{\pmf{A}^n} &=
	\frac{1}{\ln(2)(|\mset{C}_k| + \ell)^2}
	-2a \frac{\sum\limits_{i=0}^{k}\binom{n}{i}(k+1-i)}{ \left(|\mset{C}_k| + \ell\right)^3}.\label{eq:secondderivative}
	\end{align}
	The first derivative evaluates to zero only for \begin{equation*}
	  \ell_0= a\ln(2)\sum_{i=0}^{k}\binom{n}{i}(k+1-i) - |\mset{C}_k|
	\end{equation*}
	which can be negative but is larger than $-|\mset{C}_k|$.
	Evaluated at $\ell_0$, the second derivative \eqref{eq:secondderivative} is negative, hence $\ell_0$ \emph{maximizes} $\diverg{U_\mset{C}}{\pmf{A}^n}$. 
	
	We look for the integer $\hat{\ell}\in\{0,1,\dots,\binom{n}{k+1}\}$ that minimizes $\diverg{U_\mset{C}}{\pmf{A}^n}$. If $\ell_0\in [0,\binom{n}{k+1}]$, then $\diverg{U_\mset{C}}{\pmf{A}^n}$ increases on $\{0,\dots,\lfloor \ell_0\rfloor\}$ and decreases on $\{\lceil \ell_0\rceil,\dots,\binom{n}{k+1}\}$.
	If $\ell_0\notin [0,\binom{n}{k+1}]$, then $\diverg{U_\mset{C}}{\pmf{A}^n}$ is either monotonically increasing or decreasing on $\{0,1,\dots,\binom{n}{k+1}\}$. In both cases we either have $\hat{\ell}=0$ or $\hat{\ell}=\binom{n}{k+1}$.
	Thus, for codebooks of size $|\mset{C}| \in \lbrace|\mset{C}_k|,\ldots,|\mset{C}_{k+1}|\rbrace$ the \emph{minimal} divergence codebook is either $\mset{C}_k$ or $\mset{C}_{k+1}$. Since this holds for every $k$, the proof is completed.	
\end{IEEEproof}

Thus, we only consider codebooks $\mset{C}$ that are unions of type sets.
We have
\begin{equation}\label{eq:popt}
  \pletterk=\Pletterk(1) = 1-\Pletterk(0)=\frac{\sum_{i=0}^{k}\binom{n}{i}i}{\sum_{i=0}^{k}\binom{n}{i}n}
\end{equation}
which increases monotonically in $k$.
 	
\begin{lemma}\label{lem:PropertiesOfPck}
    For every positive integer $n$ and every non-negative integer $k$, $k<n/2$, we have
    \begin{equation}\label{eq:pletter_bounds_in_k}
      0 \le  \frac{k}{n} - \pletterk  \le \frac{1-k/n}{n(1-2k/n)} + \frac{1}{2n^2(1-2k/n)^2}
    \end{equation}	 	
\end{lemma}
	 
\begin{IEEEproof}
  Since $\pletterk$ is the average weight of the codewords in $\mset{C}_k$, the lower bound follows immediately.
  For the upper bound, we write
  \begin{equation}
    \pletterk = \frac{\sum_{i=0}^{k}\binom{n}{i}i}{\sum_{j=0}^{k} \binom{n}{j}n} =
    \frac12 - \frac{\sum_{i=0}^{k}\binom{n}{i}(\frac{n}{2} -i)}{\sum_{j=0}^{k} \binom{n}{j}n}.
  \end{equation}
  Using Lemma \ref{lem:SumCenterweightedBinomials}, we can simplify this to
  \begin{align}
  \pletterk&= \frac12 - \frac{\frac{k+1}{2} \binom{n}{k+1}}{\sum_{j=0}^{k} \binom{n}{j}n}=\frac12 - \frac{\frac{n-k}{2} \binom{n}{k}}{\sum_{j=0}^{k} \binom{n}{j}n} \nonumber\\
  &= \frac12 - \left(\frac12 - \frac{k}{2n} \right)\frac{ \binom{n}{k}}{\sum_{j=0}^{k} \binom{n}{j}}.
  \end{align}
  Let $q:=k/n < \frac{1}{2}$ so that $k=nq$. We insert the lower bound of Lemma \ref{lem:BoundsSumBinomial} to obtain
  \begin{align}
    \pletterk &= \frac12 - \left(\frac12 - \frac{q}{2} \right)\frac{ \binom{n}{nq}}{\sum_{j=0}^{nq} \binom{n}{j}}\notag\\
    & \geq \frac12 - \left(\frac12 - \frac{q}{2} \right)\frac{1-2q+1/n}{1-q+1/n} \left(1+\frac{1}{n(1-2q)^2}\right)\notag\\
    & \geq \frac12 - \left(\frac12 - \frac{q}{2} \right)\frac{1-2q+1/n}{1-q} \left(1+\frac{1}{n(1-2q)^2}\right)\notag\\
    & = q - \frac{1-q}{n(1-2q)} - \frac{1}{2n^2(1-2q)^2}
  \end{align}
  which establishes the upper bound of \eqref{eq:pletter_bounds_in_k}.
\end{IEEEproof}

We next show that the optimal codebook leads to a divergence that grows at least logarithmically in $n$.
Let
\begin{equation}
 \kopt=\kopt(n) := \argmin_{k\in\{0,\dots,n\}} \diverg{\unif{\mset{C}_k}}{\pmf{A}^n}
\end{equation}
for a given $n$ and a given target distribution $\pmf{A}$.
With~\eqref{eq:binaryCost} we have
 \begin{equation}
 \diverg{\unif{\Copt}}{\pmf{A}^n} =  \log_2 \frac{1}{\vert \Copt\vert} + n\entrp{\popt}+n\diverg{\popt}{p}.
%  \\ &=\diverg{P_{\Copt}}{\Pletter^n}+n\diverg{\popt}{p},
 \end{equation}
From the discussion in Section~\ref{sec:ccdm} we know that the unnormalized divergence of a CCDM grows logarithmically in $n$.
Thus, the unnormalized divergence of the optimal codebook cannot grow faster than logarithmically in $n$,
and we have $\diverg{\popt}{p}\to 0$ with $n\to\infty$.
Pinsker's inequality thus implies $|\popt-p|\to 0$.
From Lemma~\ref{lem:PropertiesOfPck} we know that $|\popt-\kopt/n|\to 0$ and hence $|\kopt/n-p|\to 0$ by the triangle inequality.

For every $k/n$, $k/n<\frac12$, we obtain the following upper bound from Lemmas \ref{Lem:ashBoundBinomial} and \ref{lem:BoundsSumBinomial}:
\begin{equation}
 \vert  \mset{C}_k \vert \leq \binom{n}{k} \frac{n-k+1}{n-2k+1} \leq \frac{2^{n\entrp{k/n}}(n-k+1)}{\sqrt{2\pi n \frac{k}{n}\frac{n-k}{n}}(n-2k+1)}.
\end{equation}
Consequently, for any codebook with $k/n<\frac12$ we have
\begin{align}
&\diverg{\unif{\mset{C}_k}}{\pmf{A}^n}\nonumber\\
&\geq  \log_2 \frac{1}{\vert \mset{C}_k\vert} + n\entrp{\pletterk}\nonumber\\
&\geq -\log_2\left( \frac{2^{n\entrp{k/n}} (n-k+1)}{\sqrt{2\pi n \frac{k}{n}\frac{n-k}{n}} (n-2k+1)}\right) + n \entrp{\pletterk}\nonumber\\
& = \frac12 \log_2 n - n(\entrp{k/n}- \entrp{\pletterk})\nonumber\\ 
&{}\quad + \frac12 \log_2\left(\frac{2\pi \frac{k}{n}(1-\frac{k}{n}) (1-2\frac{k}{n}+\frac1n)^2}{(1-\frac{k}{n}+\frac1n)^2} \right) \label{eq:diverg_lowerbound_inbetween}.
\end{align}
For small $n$, $\kopt/n$ may be greater than $1/2$; but since $\kopt/n\to p<1/2$, the above bound holds also for the optimal codebook for $n$ sufficiently large.

Now define $\epsilon(n) = \frac{1-\kopt/n}{n(1-2\kopt/n)} + \frac{1}{2n^2(1-2\kopt/n)^2}$. Lemmas~\ref{lem:asymptoticEntropyB} and~\ref{lem:PropertiesOfPck} give
\begin{align}\label{eq:entropy_Diff_optimalkn_vs_letter}
&n\entrp{\frac{\kopt}{n}}- n\entrp{\popt} \leq\nonumber\\
&\left(\frac{1-\kopt/n}{1-2\kopt/n} + \frac{1}{2n(1-2\kopt/n)^2}\right)
\log_2\frac{1-\frac{\kopt}{n}+\epsilon(n)}{\frac{\kopt}{n}-\epsilon(n)}.
\end{align}
Combining the results \eqref{eq:diverg_lowerbound_inbetween} and \eqref{eq:entropy_Diff_optimalkn_vs_letter}, we obtain a lower bound on the unnormalized divergence for $n$ sufficiently large such that $\qopt:=\kopt/n$ remains smaller than $1/2$:
\begin{align}\label{eq:divergenceEnd}
&\diverg{\unif{\mset{C}_{\kopt}}}{\pmf{A}^n} \notag\\
&\geq \frac12 \log_2 n \notag\\
&{}\quad- \left(\frac{1-\qopt}{1-2\qopt} + \frac{1}{2n(1-2\qopt)^2}\right)\log_2\frac{1-\qopt+\epsilon(n)}{\qopt-\epsilon(n)}\notag\\
&{}\quad+\frac12 \log_2\left(\frac{2\pi \qopt(1-\qopt) (1-2\qopt+\frac1n)^2}{(1-\qopt+\frac1n)^2} \right).
\end{align}

To show that the unnormalized divergence grows logarithmically in $n$, we evaluate~\eqref{eq:divergenceEnd} in the limit $n\to\infty$. Specifically, since $\qopt\to p$ and $\epsilon(n)\to 0$ with $n\to\infty$, we have
\begin{multline}
 \liminf_{n\to\infty} \left(\diverg{\unif{\mset{C}_{\kopt}}}{\pmf{A}^n} - \frac12 \log_2 n\right) \\
 \ge\frac12 \log_2\left(\frac{2\pi p(1-2p)^2}{1-p} \right) -\frac{1-p}{(1-2p)}\log_2\frac{1-p}{p}.
\end{multline}
Thus, the divergence grows at least logarithmically in $n$.

\section{Conclusion and Outlook}

We have analyzed fixed-to-fixed-length, binary-output, one-to-one DMs in terms of the unnormalized divergence between the DM output and a discrete memoryless source.
We showed that the unnormalized divergence of CCDMs grows \emph{at most} logarithmically in the output block length $n$.
For optimal DMs, the codebooks of which are unions of type sets, we showed that the unnormalized divergence grows at least logarithmically in the output block length $n$.
Thus, CCDMs perform within a constant gap of the optimal DM.
Our results suggest that fixed-to-fixed-length, binary-output, one-to-one DMs are useful in energy efficient communication schemes for DMCs, but they may fail to provide stealth.

We have reason to believe that fixed-to-fixed-length, binary-output, randomized one-to-many DMs -- which are still invertible -- may also lead to smaller unnormalized divergences. Future work shall characterize the resulting trade-off between unnormalized divergence and the randomness required for the scheme.

\section*{Acknowledgments}
We would like to thank Gerhard Kramer and Georg B\"ocherer, Technical University of Munich, for fruitful discussion and comments. The authors are indebted to Laurent Schmalen, Nokia Bell Labs, for his help in correcting Lemma~\ref{lem:PropertiesOfPck}.
The work of Patrick Schulte and Bernhard C. Geiger was supported by
the German Federal Ministry of Education and Research 
in the framework of an Alexander von Humboldt Professorship.
The work of Bernhard C. Geiger was funded by 
the Erwin Schr\"odinger Fellowship J 3765 of the Austrian Science Fund.

\bibliographystyle{IEEEtran}
\bibliography{IEEEabrv,confs-jrnls,references}

 \end{document}

%% file: preamble.tex
\usepackage{amsmath}
\usepackage{amsthm}
\usepackage{amssymb}
\usepackage{bm}
\usepackage{xspace}
\usepackage{xcolor}
\usepackage{graphicx}
\usepackage{url}
\usepackage{framed}
\usepackage{float}
\usepackage{rotating}
\usepackage{verbatim}
\usepackage{listings}
\usepackage{lscape}
\usepackage[normalem]{ulem}

\usepackage{cite}

\newcommand{\executeiffilenewer}[3]{%
\ifnum\pdfstrcmp{\pdffilemoddate{#1}}%
{\pdffilemoddate{#2}}>0%
{\immediate\write18{#3}}\fi%
}

\graphicspath{{figures/}}

\theoremstyle{plain}
\newtheorem{lemma}{Lemma}

\newcounter{algocount}
\newcounter{examplecount}

\newcommand{\bmm}{\begin{matrix}}
\newcommand{\emm}{\end{matrix}}
\newcommand{\bpm}{\begin{pmatrix}}
\newcommand{\epm}{\end{pmatrix}}

\newcommand{\bsbm}{\left[\begin{smallmatrix}}
\newcommand{\esbm}{\end{smallmatrix}\right]}
\newcommand{\bspm}{\left(\begin{smallmatrix}}
\newcommand{\espm}{\end{smallmatrix}\right)}

\newcommand{\bbm}{\begin{bmatrix}}
\newcommand{\ebm}{\end{bmatrix}}

\DeclareMathOperator*{\argmin}{argmin}

\newcommand{\mset}[1]{\mathcal{#1}}
\newcommand{\rv}[1]{\mathsf{#1}}
\newcommand{\pmf}[1]{P_{\rv{#1}}}
\newcommand{\pmfn}[2]{P_{\rv{#1}^{#2}}}

\newcommand{\entrp}[1]{\mathbb{H}\!\left( #1 \right)}

\newcommand{\diverg}[2]{\mathbb{D}\!\left( #1 \Vert #2 \right)}

%% file: diagrams/memoryless_Source.tex
\usetikzlibrary{fit}
%\usetikzlibrary{backgrounds}
%\sf
\tikzset{%
 dmsblock/.style = {draw,thick,top color = gray!1,bottom color = gray!10,rectangle,text centered,minimum height = 25,minimum width = 25},
 dsblock/.style = {draw,thick,top color = gray!1,bottom color = gray!10,rectangle,text centered,text width = 30,minimum height = 30, dash pattern=on 1pt off 4pt on 6pt off 4pt}
}
\tikzset{dashdot/.style={draw=black!50!white, line width=1pt,
                               dash pattern=on 1pt off 4pt on 6pt off 4pt,
                                inner sep=1.2mm, rectangle, rounded corners}}

\begin{tikzpicture}[scale= 0.9,auto, thick, node distance=39] %, >=triangle 45
\node (xes) {\footnotesize$\rv{B}^{m}$};
	\node [dmsblock, right of = xes, node distance=50](matcher){\footnotesize Matcher};
	\node [right of = matcher, minimum height = 40,node distance=50] (out_dev){\footnotesize$\tilde{\rv{A}}^{n}$};
    \node [dmsblock, right of = out_dev, node distance=55] (dematcher){\footnotesize Dematcher};
	\node [right of = dematcher, minimum height = 40,node distance=50] (recover){\footnotesize$\rv{B}^{m}$};
	\node [dmsblock,below of= matcher](pa){\footnotesize$P_{\rv{A}}$};
	\node [below of=out_dev, minimum height = 40](AAs){\footnotesize$\rv{A}^{n}$};
	%\begin{pgfonlayer}{background}
	\node [dashdot,fit={(matcher) (xes)}](ptilde){};
	%\end{pgfonlayer}
	
	\node[above] at (ptilde.north) {\footnotesize$P_{\mathsf{\tilde{\rv{A}}}^n}$};
	\draw[-latex] (xes) -- (matcher);
	\draw[] (matcher) -- (out_dev);
	\draw[-latex] (out_dev) -- (dematcher);
	\draw[] (dematcher) -- (recover);
	\draw[](pa) -- (AAs);
	%\path [-latex, draw, dashed](out_dev) -- node {$\approx$}(out_sim);
\end{tikzpicture}

%% file: LimitsOfFixedLength.bbl
% Generated by IEEEtran.bst, version: 1.14 (2015/08/26)
\begin{thebibliography}{10}
\providecommand{\url}[1]{#1}
\csname url@samestyle\endcsname
\providecommand{\newblock}{\relax}
\providecommand{\bibinfo}[2]{#2}
\providecommand{\BIBentrySTDinterwordspacing}{\spaceskip=0pt\relax}
\providecommand{\BIBentryALTinterwordstretchfactor}{4}
\providecommand{\BIBentryALTinterwordspacing}{\spaceskip=\fontdimen2\font plus
\BIBentryALTinterwordstretchfactor\fontdimen3\font minus
  \fontdimen4\font\relax}
\providecommand{\BIBforeignlanguage}[2]{{%
\expandafter\ifx\csname l@#1\endcsname\relax
\typeout{** WARNING: IEEEtran.bst: No hyphenation pattern has been}%
\typeout{** loaded for the language `#1'. Using the pattern for}%
\typeout{** the default language instead.}%
\else
\language=\csname l@#1\endcsname
\fi
#2}}
\providecommand{\BIBdecl}{\relax}
\BIBdecl

\bibitem{kschischang1993optimal}
F.~R. Kschischang and S.~Pasupathy, ``Optimal nonuniform signaling for
  {G}aussian channels,'' \emph{{IEEE} Trans. Inf. Theory}, vol.~39, no.~3, pp.
  913--929, 1993.

\bibitem{Ungerboeck2002}
G.~Ungerb\"ock, ``Huffman shaping,'' in \emph{Codes, Graphs, and Systems},
  R.~Blahut and R.~Koetter, Eds.\hskip 1em plus 0.5em minus 0.4em\relax
  Springer, 2002, ch.~17, pp. 299--313.

\bibitem{bocherer2011matching}
G.~B\"ocherer and R.~Mathar, ``Matching dyadic distributions to channels,'' in
  \emph{Proc. Data Compression Conf.}, Snowbird, UT, USA, 2011, pp. 23--32.

\bibitem{amjad2013fixed}
R.~A. Amjad and G.~B\"ocherer, ``Fixed-to-variable length distribution
  matching,'' in \emph{Proc. IEEE Int. Symp. Inf. Theory (ISIT)}, Istanbul,
  Turkey, 2013, pp. 1511--1515.

\bibitem{Cai2007}
N.~Cai, S.-W. Ho, and R.~Yeung, ``Probabilistic capacity and optimal coding for
  asynchronous channel,'' in \emph{Proc. IEEE Inf. Theory Workshop (ITW)}, Lake
  Tahoe, CA, USA, 2007, pp. 54--59.

\bibitem{bocherer2013arithmetic}
S.~Baur and G.~B\"ocherer, ``Arithmetic distribution matching,'' in \emph{Proc.
  Int. ITG Conf. Syst. Commun. Coding}, Hamburg, Germany, Feb. 2015, pp. 1--6.

\bibitem{amjad2013algorithms}
R.~A. Amjad, ``Algorithms for simulation of discrete memoryless sources,''
  Master's thesis, Technical University of Munich, Institute for Communications
  Engineering, 2013.

\bibitem{mondelli2014achieve}
M.~Mondelli, S.~H. Hassani, and R.~Urbanke, ``How to achieve the capacity of
  asymmetric channels,'' in \emph{Proc. Allerton Conf. Commun., Contr.,
  Comput.}, Monticello, IL, USA, Sep. 2014, pp. 789--796.

\bibitem{schulte2015constant}
P.~Schulte and G.~{B\"ocherer}, ``Constant composition distribution matching,''
  \emph{{IEEE} Trans. Inf. Theory}, vol.~62, pp. 430--434, Jan. 2016.

\bibitem{bocherer2011operating}
G.~B\"ocherer and R.~Mathar, ``Operating {LDPC} codes with zero shaping gap,''
  in \emph{Proc. IEEE Inf. Theory Workshop (ITW)}, Paraty, Brasil, 2011, pp.
  330--334.

\bibitem{bocherer2015bandwidth}
G.~B{\"o}cherer, F.~Steiner, and P.~Schulte, ``Bandwidth efficient and
  rate-matched low-density parity-check coded modulation,'' \emph{{IEEE} Trans.
  Commun.}, vol.~63, no.~12, pp. 4651--4665, Dec 2016.

\bibitem{hou2014effective}
J.~Hou and G.~Kramer, ``Effective secrecy: Reliability, confusion and
  stealth,'' in \emph{Proc. IEEE Int. Symp. Inf. Theory (ISIT)}, Jun. 2014, pp.
  601--605.

\bibitem{csiszar2004information}
I.~Csisz{\'a}r and P.~C. Shields, ``Information theory and statistics: A
  tutorial,'' \emph{Foundations and Trends® in Commun. Inf. Theory}, vol.~1,
  no.~4, pp. 417--528, 2004.

\bibitem{ash1965information}
R.~B. Ash, \emph{Information Theory}.\hskip 1em plus 0.5em minus 0.4em\relax
  New York: Dover, 1965.

\bibitem{bahadur1960approximations}
R.~R. Bahadur, ``Some approximations to the binomial distribution function,''
  \emph{Ann. Math. Statistics}, pp. 43--54, 1960.

\bibitem{graham1989concrete}
R.~L. Graham, D.~E. Knuth, and O.~Patashnik, \emph{Concrete Mathematics},
  2nd~ed., 1989.

\bibitem{bocherer2014optimal}
G.~B\"ocherer and B.~C. Geiger, ``Optimal quantization for distribution
  synthesis,'' \emph{{IEEE} Trans. Inf. Theory}, vol.~62, no.~11, pp.
  6162--6172, Nov. 2016, preprint available: {\tt arXiv:1307.6843 [cs.IT]}.

\end{thebibliography}
